\documentclass[twocolumn,prb,showpacs,amssymb,superscriptaddress,nofootinbib]{revtex4-1}

\usepackage{graphicx,hyperref}
\usepackage{amsmath}
\usepackage{amsfonts}
\usepackage{epsfig} 
\usepackage{color}
\usepackage{bbm}
\usepackage[normalem]{ulem}

\usepackage{amsthm}

\newcommand{\defeq}{\mathrel{\mathop:}=}

\begin{document}
\title{Bethe-Boltzmann Hydrodynamics and Spin Transport in the XXZ Chain}

\author{Vir B. Bulchandani}
\affiliation{Department of Physics, University of California, Berkeley, Berkeley CA 94720, USA}

\author{Romain Vasseur}
\affiliation{Department of Physics, University of California, Berkeley, Berkeley CA 94720, USA}
\affiliation{Materials Sciences Division, Lawrence Berkeley National Laboratory, Berkeley CA 94720, USA}
\affiliation{Department of Physics, University of Massachusetts, Amherst, MA 01003, USA}

\author{Christoph Karrasch}
\affiliation{Dahlem Center for Complex Quantum Systems and Fachbereich Physik, Freie Universit\"at Berlin, 14195 Berlin, Germany}

\author{Joel E. Moore}
\affiliation{Department of Physics, University of California, Berkeley, Berkeley CA 94720, USA}
\affiliation{Materials Sciences Division, Lawrence Berkeley National Laboratory, Berkeley CA 94720, USA}

\date{\today}

\begin{abstract}

Quantum integrable systems, such as the interacting Bose gas in one dimension and the XXZ quantum spin chain, have an extensive number of local conserved quantities that endow them with exotic thermalization and transport properties. We discuss recently introduced hydrodynamic approaches for such integrable systems from the viewpoint of kinetic theory and extend the previous works by proposing a numerical scheme to solve the hydrodynamic equations for finite times and arbitrary locally equilibrated initial conditions. We then discuss how such methods can be applied to describe non-equilibrium steady states involving ballistic heat and spin currents. In particular, we show that the spin Drude weight in the XXZ chain, previously accessible only by rigorous techniques of limited scope or controversial thermodynamic Bethe ansatz arguments, may be evaluated from hydrodynamics in very good agreement with density-matrix renormalization group calculations. 
\end{abstract}
\maketitle
\section{Introduction}

The study of many-body quantum systems far away from equilibrium conditions poses a considerable challenge for theory, even for quantum integrable systems whose equilibrium properties may be computed exactly. Such systems, which include the Heisenberg antiferromagnet and the Lieb-Liniger gas in one dimension, possess an extensive number of conserved quantities, which prevent them from thermalizing like generic ergodic systems and lead to dissipation-less transport properties. Under unitary evolution, the local properties of these systems are believed to tend to a generalized Gibbs ensemble (GGE)~\cite{Rigol:2008kq,PhysRevLett.100.100601} at long times, containing in principle {\it all} the independent conserved quantities, not just particle number and energy as in the standard Gibbs ensemble. The rest of the system acts as an ``unusual bath''.

However, this convergence can be rather subtle, as the GGEs constructed using the standard conserved quantities of the spin-$1/2$ XXZ chain~\cite{1742-5468-2013-07-P07003,1742-5468-2013-07-P07012,PhysRevB.89.125101} were shown~\cite{PhysRevLett.113.117202,PhysRevLett.113.117203} to fail to reproduce the correct steady states obtained either numerically or using the so-called quench action method~\cite{PhysRevLett.110.257203}. This paradox can only be resolved~\cite{PhysRevLett.115.157201} by taking into account ``hidden'' (non-standard) quasi-local conserved quantities~\cite{prosenxxz,PhysRevLett.111.057203,Prosen20141177,1742-5468-2014-9-P09037,PhysRevLett.115.120601} in the GGE.  More recently, the nature of these new conserved quantities and their relation to the pseudo-momentum distributions used in equilibrium thermodynamic Bethe ansatz calculations was clarified~\cite{1742-5468-2016-6-063101,2016arXiv161006911I,2016arXiv161207265D}.

Nevertheless, when combined with the infinite number of Lagrange multipliers which must be fixed from the initial state, this complication makes the GGE approach quite cumbersome for the study of interesting non-equilibrium dynamics in integrable models, particularly those arising from spatially non-uniform states. Moreover, generalized Gibbs ensembles are by definition restricted to describe steady-states, making a general approach able to deal with non-equilibrium finite-time dynamics desirable. Very recently, a more practical hydrodynamic approach based on a semi-classical quasi-particle picture was introduced~\cite{Doyon,Fagotti}, and conjectured to yield exact results for the long-time scaling limit reached from the ``two-reservoir quench,'' an initial condition of two semi-infinite reservoirs connected at the origin. One way to understand these approaches is that they reflect an equivalence between the hydrodynamical and Boltzmann-type descriptions of integrable models.

In standard statistical mechanics, the Boltzmann equation involves the full one-particle distribution function, i.e., a function of momentum at each point in space and time. Standard hydrodynamics contains considerably less information, as only three quantities survive at each point in space and time: the local particle density, momentum density, and energy density. For integrable models, there is a more fundamental relationship between the distribution function $\rho(x,t,k)$ over pseudo-momentum $k$ (the analogue of ordinary momentum for integrable models) and the full set of conserved quantities. This is most easily seen in the Lieb-Liniger model, where the conserved quantities are just moments of the pseudo-momentum distribution~\cite{Korepin}. It is much less clear that it holds for the XXZ model, whose conserved quantities have a rather subtle structure (see Refs.~\onlinecite{1742-5468-2016-6-063101,2016arXiv161006911I,2016arXiv161207265D} for recent developments), but we show in this paper that a hydrodynamical description is successful even for observables that are sensitive to the newly discovered quasi-local charges. This result is consistent with recent work demonstrating the equivalence of the GGE and thermodynamic Bethe ansatz (TBA) pictures of equilibrium states in the XXZ model~\cite{2016arXiv161207265D}, although some subtleties of interpretation still remain\cite{Pozsgay}.

The first derivations of the recent hydrodynamic formalism were based upon the observation that making a local-density-type approximation for \textit{all} local conserved charges implies a conservation law at the level of the local pseudo-momentum distribution. The latter conservation law, previously obtained in Refs. \onlinecite{Doyon, Fagotti}, can alternatively be viewed as a dissipationless Boltzmann equation for the pseudo-momentum distribution, where the effect of collisions is to dress the bare velocity of particle-type excitations. It is precisely this kinetic equation that was solved in Refs. \onlinecite{Doyon,Fagotti}, rather than an infinite set of hydrodynamic equations, and in recognition of its crucial role within the hydrodynamic formalism, we call this equation the ``Bethe-Boltzmann equation''. Our terminology seeks to emphasize that this equation can be understood phenomenologically as a Boltzmann equation whose collision term captures the Wigner time-delay \cite{Wigner} due to Bethe ansatz phase shifts (this is one way to interpret the velocity dressing equation, Eq. \eqref{vdressedq}).

From this perspective, the Bethe-Boltzmann equation emerges as the quantum analogue of an established kinetic theory of classical soliton gases\cite{Percus, Zakharov,Boldrighini1983,El1,El2}. Our viewpoint is also consistent with a recently obtained ``molecular dynamics'' realization of this equation\cite{DoyonSoliton}. This viewpoint is useful because it provides some additional physical justification for applying this equation at finite- time and length- scales, which, although admittedly less rigorous than the discussion of quasi-stationary states in Refs. \onlinecite{Doyon, Fagotti}, is closer in spirit to the semi-classical theory of quasiparticle transport that underpins more conventional solid-state physics\cite{AshMerm}.

This relatively simple picture of the Bethe-Boltzmann equation raises the question of whether predictions obtained from the hydrodynamic formalism are in fact correct. Fortunately, there exist a plethora of predictions for non-equilibrium evolution of quantum integrable models from a range of initial conditions, against which hydrodynamics can be compared. Among these, the two-reservoir quench mentioned above has had an enduring popularity; this set-up consists of two half-infinite systems prepared in thermal equilibrium with different temperatures and chemical potentials (or more generally, two half-infinite homogeneous reservoirs), joined together at time $t=0$ and allowed to evolve unitarily for $t>0$ according to the Schr\"odinger equation~\cite{sotiriadiscardy,bernarddoyon,karraschilanmoore,Doyon2015190,PhysRevB.88.134301,Bhaseen:2015aa,BernardDoyonReview,VKM,PhysRevB.90.161101,PhysRevB.93.205121,1742-5468-2016-6-064005,1742-5468-2016-6-064010,2017arXiv170106620L}. In light of generalized hydrodynamics, this gives rise to the intriguing possibility of deriving aspects of non-equilibrium quantum transport from an essentially \textit{classical} equation.  Non-equilibrium thermal transport in the XXZ spin chain from hydrodynamics was studied in detail in Ref.~\onlinecite{Fagotti} and was thoroughly compared to matrix-product state numerics~\cite{white92,schollwoeck}.

The case of spin transport is somewhat more complex theoretically, even at the level of linear response, as the presence of ballistic spin currents requires the existence of quasi-local conserved quantities~\cite{prosenxxz} going beyond the standard local conserved quantities of the XXZ spin chain. For conventional (linear response) transport, the ballistic component of the spin current is characterized by the spin Drude weight, which measures the degree of divergence of the zero-frequency spin conductivity. The Bethe ansatz calculation of the spin Drude weight~\cite{PhysRevLett.82.1764,doi:10.1143/JPSJS.74S.181} has attracted a lot of attention in the past, and remains controversial. In fact, before the discovery of the quasi-local conserved quantities mentioned above, it was even debated whether this Drude weight was non-zero at finite temperature (see {\it e.g.} Refs.~\onlinecite{PhysRevB.68.134436,sirker:2010,PhysRevB.83.035115,karraschdrude,PhysRevB.87.245128} and references therein).
In this paper, we devise a method to compute the spin Drude weight from the Bethe-Boltzmann equation, and show that the value of the resulting Drude weight is compatible with known exact results and with density-matrix renormalization group (DMRG) calculations. This is remarkable because it indicates that the non-equilibrium steady-state predicted by hydrodynamics takes all quasi-local conserved charges into account (see also Ref. ~\onlinecite{2016arXiv161207265D}), even though such charges can effectively be ignored at the level of the Bethe-Boltzmann equation, which could have been deduced on phenomenological grounds long before the discovery of quasi-local conservation laws.

We proceed as follows. In Section \ref{SecII}, we summarize the background from Refs.~\onlinecite{Doyon, Fagotti} needed to formulate our main results on the spin Drude weight. We also discuss how the Bethe-Boltzmann equation can be viewed as a phenomenological extension of older kinetic theories for classical soliton gases to quantum integrable models (see also \onlinecite{DoyonSoliton}), which provides additional physical motivation for applications of the hydrodynamic approach at finite time- and length- scales. We mostly defer the detailed calculation of this finite-time hydrodynamics in physically relevant examples for a subsequent publication~\cite{Unpub}, except for one example, a thermal expansion in the XXZ chain, in Figure \ref{figFiniteTime} below. This extends the analysis in previous works\cite{Doyon,Fagotti} by providing the first numerical checks of the hydrodynamic approach beyond scale-invariant steady-states. In Section \ref{SecIII} we present our main results, on the evaluation of the spin Drude weight $D_s(T)$.

\section{The Bethe-Boltzmann Equation}
\label{SecII}
The reader familiar with Refs. 18 and 19 can skip Section \ref{SecII} without loss of continuity. Here, we summarize the background from Refs. 18 and 19 that is necessary for formulating our main results on the spin Drude weight, in Section \ref{SecIII}. We also discuss the intimate connection between the Bethe-Boltzmann equation and various existing kinetic equations describing the dynamics of classical soliton gases\cite{Percus, Zakharov,Boldrighini1983,El1,El2}.
\subsection{Motivation}
\label{SecIIA}
To provide some intuition, we first sketch the Bethe-Boltzmann formalism for the one-dimensional Bose gas with delta-function interactions. For $N$ bosons on a line, with interaction strength $c$, the Hamiltonian may be written as
\begin{equation}
\hat{H} = - \sum_{j=1}^N \partial_j^2 + \sum_{j<k}2c\delta(x_j-x_k).
\end{equation}
This model, also called the Lieb-Liniger gas, is the simplest non-trivial integrable model, and amongst the entire class of such models has the merit of being the most relevant to experimental physics. The physics of quasiparticle excitations in this model, as obtained from TBA, is summarized in Appendix \ref{AppB}. Let us now consider a Lieb-Liniger gas on a line of length $L$, consisting of $N$ particles. In the thermodynamic limit as $N,L\to \infty$, we assume that the system may be characterized by a local density of occupied pseudo-momenta, $\rho(x,t,k)$, giving rise to a locally varying particle density
\begin{equation}
n(x,t) = \int_{-\infty}^{\infty} dk \, \rho(x,t,k).
\end{equation}
Physically speaking, this amounts to coarse-graining our line into cells of length $l \ll L$, such that on each cell, the gas lies in a macrostate fixed by the average particle and hole densities over that cell. We additionally postulate that ``occupied quantum numbers are locally conserved''. This implies a local conservation law of the form
\begin{equation}
\partial_t \rho(x,t,k) + \partial_x j(x,t,k) = 0,
\end{equation}
for some current $j(x,t,k)$ to be determined. To obtain a specific form for $j$, observe that a physically natural velocity scale for the transport of quantum numbers is given by the quasiparticle velocity $v[\rho](x,t,k)$ in each cell, which generally depends on the local occupation number $\{\rho(x,t,k):k\in \mathbb{R}\}$. This yields
\begin{equation}
\partial_t \rho(x,t,k) + \partial_x(\rho(x,t,k)v[\rho](x,t,k)) = 0,
\label{eqBB}
\end{equation}
which we shall henceforth refer to as the \textit{Bethe-Boltzmann equation}. This equation has the structure of a conservation law for the local pseudo-momentum distribution $\rho$, which is how it was first introduced in Refs. \onlinecite{Doyon,Fagotti}, where it was derived from an infinite set of ``generalized hydrodynamic'' equations. However, Eq. \eqref{eqBB} can be obtained directly by the above phenomenological arguments, and from this viewpoint, defines a Boltzmann-type equation for the dissipationless transport of quasiparticles. By varying the number of local densities appropriately, an equation of this type may be formulated for \textit{any} quantum integrable model. We now discuss this formalism in more detail, and in particular, show how it fits into an existing kinetic theory for solitons of classically integrable PDEs.

\subsection{Relation with the Kinetic Theory of Solitons of Classically Integrable PDEs}
\label{SecIIB}
It has long been known that there exists a class of non-linear partial differentiable equations, called ``integrable'', which admit stably propagating ``solitonic'' solutions. For example, the KdV equation
\begin{equation}
\phi_t = 6 \phi \phi_x - \phi_{xxx},
\end{equation}
is of this type. In 1971, shortly after the discovery of the classical inverse scattering method for solving such equations, Zakharov\cite{Zakharov} considered the kinetic theory of their solitons and proposed an integro-differential equation for a gas of KdV solitons in the dilute limit, of the form
\begin{equation}
\partial_t \rho(x,t,\lambda) + \partial_x(\rho(x,t,\lambda)v[\rho](x,t,\lambda)) = 0,
\label{BBclass}
\end{equation}
where $\rho(x,t,\lambda)$ denotes the local density of solitons with spectral parameter $\lambda$ and $v$ their effective velocity, given as a certain functional linear in $\rho$. Several years later, an extension of this equation beyond the dilute limit was obtained from an infinite-genus limit of the Whitham equations\cite{El1}. It was found that the equation \eqref{BBclass} held in general, provided that Zakharov's explicit formula for $v$ was replaced by an implicit integral equation 
\begin{equation}
v(\lambda) = v_0(\lambda) + \int_{-\infty}^{\infty} d\mu \, \Delta x(\lambda,\mu) \rho(\mu)[v(\lambda)-v(\mu)],
\label{vdressed}
\end{equation}
with $v_0(\lambda)$ the velocity of a single soliton with spectral parameter $\lambda$ and $\Delta x(\lambda,\mu)$ the asymptotic position shift after a collision of two KdV solitons with spectral parameters $\lambda$ and $\mu$. A similar formula was obtained rigorously in the hydrodynamic limit of a classical hard-rod gas\cite{Boldrighini1983}. Despite the mathematical complexities of deriving Eq. \eqref{vdressed}, whether for classical solitons or hard rods, its physical interpretation in terms of two-body phase shifts is straightforward. This interpretation led El and Kamchatnov to propose an equation of the form \eqref{vdressed} for arbitrary soliton gases with two-body elastic scattering, arising from a given classically integrable PDE\cite{El2}. More recently, the system of equations \eqref{BBclass}, \eqref{vdressed} was obtained in the context of Euler-scale hydrodynamics for quantum integrable systems \cite{Doyon, Fagotti}. In particular, one can show that in the quantum setting, the dressed velocities satisfy the integral equation\cite{Doyon}
\begin{equation}
v(\lambda) = v_0(\lambda) + \frac{1}{p'(\lambda)}\int_{-\infty}^{\infty} d\mu \, \varphi'(\lambda-\mu) \rho(\mu)[v(\lambda)-v(\mu)],
\label{vdressedq}
\end{equation}
where $p(\lambda)$ denotes the bare quasiparticle momentum, $e(\lambda)$ the bare energy and $v_0(\lambda)=e'(\lambda)/p'(\lambda)$ the bare group velocity. Also, the differential scattering phase is given in terms of the two-particle S-matrix by
\begin{equation}
\varphi'(\lambda-\mu) = i \frac{d}{d\lambda}\ln{S(\lambda-\mu)}.
\end{equation}
Upon comparing equations \eqref{vdressed} and \eqref{vdressedq}, it is clear that the kinetic theory for quantum solitons with differential scattering phase $\varphi'$ may be expressed as a kinetic theory for classical solitons, upon making the identification
\begin{equation}
\Delta x(\lambda,\mu) = \frac{1}{p'(\lambda)}\varphi'(\lambda-\mu).
\label{fund}
\end{equation}
This equivalence was also obtained in Ref. \onlinecite{DoyonSoliton}, independently of the existing theory in Ref. \onlinecite{El2}. Thus, although the original derivation of the equations \eqref{BBclass}, \eqref{vdressedq} in the quantum setting made use of an infinite system of Euler equations \cite{Doyon, Fagotti}, these equations are intimately related with a kinetic theory framework for solitons of classically integrable PDEs that is many decades older. 

It is worth noting that at present, both the hydrodynamic and kinetic theory viewpoints for quantum integrable models are approximate to the same degree; on the hydrodynamic side, the approximation lies in the ``Euler-scale'' assumption that diffusive terms are negligible while on the kinetic theory side, the approximation lies in neglecting all higher-order collision terms which could lead to entropy generation. Although it remains to be seen which viewpoint is better suited to incorporating higher-order effects, it appears that all of the proposed extensions of the hydrodynamic formalism for quantum integrable models that take into account new physics, such as external potentials \cite{DoyonII} and collision terms \cite{DoyonSpohn} (and indeed the recent analogy with classical solitons\cite{DoyonSoliton}), fall naturally into a Boltzmann paradigm rather than a hydrodynamic one.

A natural question to ask is whether there exist gases of solitons arising from classically integrable PDEs whose kinetic theory is captured by the Bethe-Boltzmann equation for some quantum integrable system. For example, there is a well-known mapping\cite{NLSE} between hole-type excitations of the Lieb-Liniger gas and dark solitons of the defocusing non-linear Schr{\"o}dinger equation in the weak-coupling limit, $c \to 0^+$. We have found that the position shift resulting from a collision of two dark solitons\cite{Faddeev} coincides with an expression for the inverse of the Lieb-Liniger kernel obtained by Gaudin in the weak-coupling limit\cite{Gaudin}.

\subsection{The Bethe-Boltzmann Equation}
\label{SecIIC}
\subsubsection{Formulation}
We now return to the precise formulation of the Bethe-Boltzmann equation~\eqref{eqBB} for the Lieb-Liniger model. Although the integral equation Eq. \eqref{vdressed} is helpful for developing the analogy with kinetic theory, it is easier in practice to use the following explicit formula for $v$. Recall that in TBA, the group velocity of dressed excitations on a given equilibrium state with Fermi factors $\{\theta(k):k \in \mathbb{R}\}$ for each pseudo-momentum $k$, is given by
\begin{equation}
v(k) = \frac{(\hat{1}+\hat{K}\hat{\vartheta})^{-1}[k'](k)}{(\hat{1}+\hat{K}\hat{\vartheta})^{-1}[1](k)}.
\label{eqVel}
\end{equation}
The manipulations required to derive this equation are summarized in Appendix B. In order to be able to use this result in the Bethe-Boltzmann formalism, we must impose an additional assumption on the local pseudo-momentum density $\rho(x,t,k)$. In particular, we need to assume that each microstate corresponding to the set of local occupation numbers $\{\rho(x,t,k) : k \in \mathbb{R}\}$ defines an eigenstate of the Lieb-Liniger Hamiltonian. Thus we demand that the \textit{local Bethe equation}
\begin{equation}
\frac{\rho(x,t,k)}{\vartheta(x,t,k)} + \int_{-\infty}^{\infty} dk'\, \mathcal{K}(k,k')\rho(x,t,k') = \frac{1}{2\pi},
\label{eqlocalBA}
\end{equation}
holds at every point; this may be taken as a definition of the \textit{local Fermi factor} $\vartheta(x,t,k)$, which in turn yields the \textit{local quasiparticle velocity}, 
\begin{equation}
v(x,t,k) = \frac{(\hat{1}+\hat{K}\hat{\vartheta}(x,t))^{-1}[k'](k)}{(\hat{1}+\hat{K}\hat{\vartheta}(x,t))^{-1}[1](k)}.
\end{equation}
To summarize, the Bethe-Boltzmann equation is shorthand for the hierarchy of equations
\begin{align}
\nonumber \partial_t \rho(x,t,k) + \partial_x (\rho(x,t,k)v(x,t,k)) &= 0 ,\\
\nonumber  \frac{2\pi\rho(x,t,k)}{1-2\pi\hat{K}[\rho(x,t,k')](k)} &= \vartheta(x,t,k), \\
\frac{(\hat{1}+\hat{K}\hat{\vartheta}(x,t))^{-1}[k'](k)}{(\hat{1}+\hat{K}\hat{\vartheta}(x,t))^{-1}[1](k)} &= v(x,t,k),
\label{loctba}
\end{align}
which together comprise a conservation law with self-consistently determined velocity. We can write this schematically as
\begin{equation}
\partial_t\rho + \partial_x(\rho v[\rho])=0.
\end{equation}
This turns out not to be the most useful form of the Bethe-Boltzmann equation. As in the previous works~\cite{Doyon,Fagotti}, we find it more convenient to view $\vartheta$ as the fundamental degree of freedom rather than $\rho$. This is possible because the local Bethe equation~\eqref{eqlocalBA} allows one to express $\rho$ and $\rho v$ as functionals of the local Fermi factor $\vartheta$, namely
\begin{align}
\nonumber \rho[\hat{\vartheta}(x,t)](k) &=\frac{1}{2\pi} (\hat{\vartheta}(x,t)^{-1} + \hat{K})^{-1}[1](k) \\
(\rho v)[\hat{\vartheta}(x,t)](k) &= \frac{1}{2\pi}(\hat{\vartheta}(x,t)^{-1} + \hat{K})^{-1}[k'](k).
\end{align}
Upon substituting these expressions into the Bethe-Boltzmann equation, a surprising simplification occurs, and a conservation law for $\rho$ is replaced by a simpler advection equation for $\vartheta$. One finds that
\begin{align}
\nonumber \partial_t \vartheta(x,t,k) + v[\hat{\vartheta}(x,t)](k)\partial_x \vartheta(x,t,k) &= 0, \\
\frac{(\hat{1} + \hat{K}\hat{\vartheta}(x,t))^{-1}[k'](k)}{(\hat{1} + \hat{K}\hat{\vartheta}(x,t))^{-1}[1](k)} &= v[\hat{\vartheta}(x,t)](k),
\end{align}
or schematically,
\begin{equation}
\partial_t \vartheta + v[\vartheta]\partial_x \vartheta = 0.
\label{eqBBadv}
\end{equation}
This is an example of a \textit{quasilinear advection equation}. In general, the time evolution of such equations rapidly leads to shock formation, even for smooth initial conditions, and without additional assumptions the initial value problem is ill-posed. For two-reservoir initial conditions, this issue was circumvented \cite{Fagotti,Doyon} by imposing a certain self-consistent ansatz, which for TBA-based hydrodynamics picks out a unique weak solution $\vartheta(x,t,k)$.

Before the present work, the equation \eqref{eqBBadv} had only been verified numerically for long-time steady states of scale-invariant quenches, a regime in which the validity of hydrodynamics can be argued using Bethe-ansatz techniques\cite{Doyon,Fagotti}. Although the possibility of applications to finite-time dynamics had been raised in previous works\cite{Doyon,DoyonII}, there had been no means to actually solve the hydrodynamic equations in the finite-time regime. 

In order to address this shortcoming, we have developed a numerical scheme to solve Eq. \eqref{eqBBadv} at finite times for evolution from arbitrary smooth, locally equilibrated initial conditions, whose details are given in Appendix \ref{SecAppC}. The simplest implementation of this scheme yields a self-consistent ansatz that generalizes the two-reservoir case. We were able to obtain physically reasonable predictions for a range of models and initial conditions, which are mostly deferred to a companion paper~\cite{Unpub} except for one example, a thermal expansion in the XXZ model. As discussed in Ref. \onlinecite{Doyon}, we expect on general physical grounds that Eq. \eqref{eqBBadv} should be accurate at time-scales which are long compared to the time-scale of local thermalization and at length scales which are long compared to the length scale over which the system is locally in thermal equilibrium, even though rigorous justification for applications beyond the scaling limit is lacking at present.

\subsection{The Two-Reservoir Quench}
Let us now outline the ansatz for non-equilibrium steady states of the two-reservoir quench already presented in Refs.~\onlinecite{Doyon, Fagotti}. Thus consider solving the initial value problem 
\begin{align}
\nonumber \partial_t \vartheta(x,t,k) + v[\hat{\vartheta}](k)\partial_x \vartheta(x,t,k) &= 0, \\
\vartheta(x,0,k) &= \phi(x,k).
\label{IVPmt}
\end{align}
for $t>0$, with discontinuous initial conditions
\begin{equation}
\phi(x,k) = \theta_L(k)H(-x)+ \theta_R(k)H(x),
\end{equation}
where $\theta_L$ and $\theta_R$ denote the Fermi factors for initial equilibrium states with temperatures and chemical potentials $\{T_L,\mu_L\}$ and $\{T_R, \mu_R\}$ respectively, as given by the Yang-Yang equations~\cite{YY}, and $H$ denotes the Heaviside step function. By analogy with the solution by characteristics for the Burgers equation, one can write down the ansatz
\begin{equation}
\vartheta(x,t,k) = \phi(x-v[\hat{\vartheta}](k)t,k),
\label{scinfmt}
\end{equation}
which yields\footnote{This only works for two reservoirs due to (i) scale invariance and (ii) the fact that TBA-based hydrodynamics is linearly degenerate, in the sense of Lax\cite{Lax}. Otherwise, existence and uniqueness of solutions to this ansatz are not guaranteed, even if one assumes monotonicity of $v[\vartheta](k)$ in $k$.}
\begin{align}
\nonumber \vartheta(x,t,k) &= \theta_L(k)H(v[\hat{\vartheta}(x,t)](k)t-x) \\
&+\theta_R(k)H(x-v[\hat{\vartheta}(x,t)](k)t).
\label{eq35}
\end{align}
In the special case that $v[\hat{\vartheta}(x,t)](k)$ is monotonic in $k$, we can write this as a step-function of the wavenumber $k$, as was done in Ref.~\onlinecite{Doyon}. To see this, suppose that for fixed $x,t$, the function  $v[\hat{\vartheta}(x,t)](k)$ is monotonic in $k$. Then the equation
\begin{equation}
v[\hat{\vartheta}(x,t)](k)t-x = 0,
\end{equation}
has a unique solution, $k^*(x,t)$, such that
\begin{equation}
v[\hat{\vartheta}(x,t)](k^*(x,t)) = x/t.
\end{equation}
Thus for example, if $v[\hat{\vartheta}(x,t)](k)$ increases with $k$, we can write
\begin{equation}
\vartheta(x,t,k) = \theta_L(k)H(k-k^*(x,t))+ \theta_R(k)H(k^*(x,t)-k).
\label{eqIt}
\end{equation}
As they stand, equations~\eqref{eq35} and~\eqref{eqIt} both appear intractable. However, when solving for long time steady-states, we can exploit the crucial property of \textit{scale-invariance}. In particular, at long times we may suppose that $\vartheta(x,t,k)$ depends on position and time via their ratio $\zeta = x/t$ alone. In~\eqref{eq35}, this yields
\begin{align}
\vartheta(\zeta,k) = &\theta_L(k)H(v[\hat{\vartheta}(\zeta)](k)-\zeta])+ \theta_R(k)H(\zeta-v[\hat{\vartheta}(\zeta)](k)),
\end{align}
which is essentially equation (16) of Ref.~\onlinecite{Fagotti}. Assuming that $v$ increases with $k$, this may be recast as the pair of self-consistent equations
\begin{align}
\nonumber v[\hat{\vartheta}(\zeta)](k^*(\zeta)) &= \zeta \\
\vartheta(\zeta,k) &= \theta_L(k)H(k-k^*(\zeta))+ \theta_R(k)H(k^*(\zeta)-k),
\end{align}
which is essentially equation (35) of Ref.~\onlinecite{Doyon}. This form is particularly amenable to iterative solution.

\subsection{Hydrodynamic Charges and Currents}
\label{secHydroChargesCurrentsLL}
In order to make contact with direct numerical simulations of the two-temperature quench, we must develop hydrodynamic expressions for local charges and currents of the model. Therefore suppose that $\mathbf{Q}$ is a conserved charge operator of the model, with single-particle eigenvalue $q(k)$. Then the total charge carried by a Bethe wavefunction with limiting density of states $\rho(k)$ is given by
\begin{equation}
 \langle Q \rangle = \sum_{j} q(k_j) \sim L \int_{-\infty}^{\infty} dk\,\rho(k)q(k),
\end{equation}
in the thermodynamic limit. Whilst this is a standard result, the surprising claim of Ref.~\onlinecite{Fagotti} (proven in the case of Lorentz or Galilean-invariant theories in Ref.~\onlinecite{Doyon}) is that the local \textit{currents} associated with such charges may also be written in terms of the local density of states, via the formula
\begin{equation}
\langle J \rangle \sim L \int_{-\infty}^{\infty} dk\,\rho(k)q(k)v(k),
\end{equation}
where $v(k)$ denotes the quasi-particle velocity. In the hydrodynamic approximation, this allows us to write down expressions for the local charge and current densities associated with the operator $\bold{Q}$, namely
\begin{align}
\langle q \rangle (x,t) &= \int_{-\infty}^{\infty} dk\,\rho(x,t,k)q(k), \\
\langle j \rangle (x,t) &= \int_{-\infty}^{\infty} dk\,\rho(x,t,k)q(k)v(x,t,k).
\end{align}
Given these definitions, the Bethe-Boltzmann equation immediately implies local conservation of charge, in the form
\begin{equation}
\partial_t \langle q \rangle (x,t) + \partial_x \langle j \rangle (x,t) = 0.
\label{eqconservation}
\end{equation}
The equation~\eqref{eqconservation}, together with a ``completeness property'' of local conserved charges, is used to derive the Bethe-Boltzmann equation by both Ref.~\onlinecite{Doyon} and Ref.~\onlinecite{Fagotti}. However, attempting to derive the Bethe-Boltzmann equation in this way at finite length- and time- scales ends up being equivalent to the phenomenological derivation we provide in Section \ref{SecIIA}, as one must \textit{a priori} make a local-density-type approximation for $\rho$, whose rigorous justification for non-stationary states does not yet exist.

\section{Linear Response and Drude Weights in the spin-$1/2$ XXZ Chain}
\label{SecIII}
Now that we are equipped with the Bethe-Boltzmann or generalized hydrodynamics framework, we illustrate how it can be applied to study energy and spin transport in the spin-$1/2$ XXZ spin chain, and compare our results to density-matrix renormalization group calculations. Even near equilibrium (linear response), we argue that this framework allows one to compute transport quantities that were previously inaccessible, including the spin Drude weight at arbitrary temperature. 
\subsection{Bethe-Boltzmann Formalism for the XXZ Chain}
Recall that the Hamiltonian for the spin-$1/2$ XXZ chain on $N$ sites in an external field $h$ is given by
\begin{equation}
H = J \sum_{j=1}^{N-1} S^x_{j}S^x_{j+1}+S^y_{j}S^y_{j+1} + \Delta S^z_{j}S^z_{j+1} + 2h\sum_{j=1}^N S^z_{j}.
\label{XXZham}
\end{equation}
Here, we take periodic boundary conditions $S_N \equiv S_{N+1}$, set the coupling to $J=1$, and parameterize the anisotropy of the theory by $\Delta = \cos \gamma$.  We assume in the following that $-1 < \Delta < 1$; the behavior outside this regime is mentioned briefly in the final Discussion.  The Bethe-Boltzmann equation for the Hamiltonian~\eqref{XXZham} is discussed in detail in Ref.~\onlinecite{Fagotti}. The derivation proceeds almost exactly as for the Lieb-Liniger gas, except that one must now account for the ``strings'' of bound states appearing in the thermodynamic limit. We define a \textit{string of type $j$} to be a an ordered pair, $(n_j,v_j)$, where $n_j$ is the number of spin-flips comprising the string and $v_j$ is its parity, and assume that there are $N_t$ string types in total. One can show that their dressed velocities are given by the formula
\begin{equation}
v_j(\lambda) =  \frac{1}{2\pi}\frac{(\hat{\sigma}+\hat{T}\hat{\vec{\vartheta}})^{-1}[-A\vec{a}']_j}{(\hat{\sigma}+\hat{T}\hat{\vec{\vartheta}})^{-1}[\vec{a}])_j}(\lambda),
\label{eqvXXZ}
\end{equation}
whose motivation is sketched in Appendix \ref{AppB}. Since there are now multiple branches of quasiparticle excitations, we must postulate multiple Bethe-Boltzmann equations; in abridged form, these read
\begin{equation}
\partial_t\rho_j + \partial_x(\rho_j v_j[\vec{\rho}])=0,\quad j = 1,2,\ldots N_t,
\end{equation}
with advection formulation
\begin{align}
\partial_t \vartheta_j +  v_j[\vec{\vartheta}]\partial_x \vartheta_j = 0, \quad j = 1,2,\ldots N_t,
\end{align}
where the $v_j$ are given by ~\eqref{eqvXXZ}. For steady-states of the two-reservoir quench, the latter may be solved formally to yield the $2N_t$ coupled equations
\begin{align}
\nonumber v_j[\hat{\vec{\vartheta}}(\zeta)](\lambda_j^*(\zeta)) &= \zeta, \\
\vartheta_j(\zeta,\lambda) &= \theta_{Lj}(k)H(\lambda-\lambda_j^*(\zeta))+ \theta_{Rj}(k)H(\lambda^*(\zeta)-\lambda),
\label{eqNum}
\end{align}
where we again assumed that the $v_j$ were monotonic in $\lambda$.

\subsection{Linear Response}

We now turn to conventional (linear response) quantum transport in integrable systems. Linear response transport coefficients are given by the Kubo formula, which relates the conductivity at zero frequency to the integral over equilibrium dynamical correlation functions describing the return to equilibrium of a spontaneous fluctuation. In an integrable system that does not thermalize in a conventional way, energy or spin (charge) currents may not be able to relax if they have a non-zero overlap with conserved quantities, leading to a divergent zero-frequency conductivity and dissipation-less transport~\cite{PhysRevB.55.11029}. The degree of divergence of the DC conductivity may be characterized by considering the conductivity at finite frequency and defining the Drude weight $D(T)$ such that
\begin{equation}
\sigma(\omega) = \pi D(T) \delta(\omega) + \sigma_{\rm regular} (\omega),
\end{equation}
 with $D(T)$ given by the long-time behavior of the equilibrium dynamical correlation function
\begin{equation}\label{drude_lr}
D(T) = \beta \lim_{t \to \infty} \frac{\langle J(t) J(0) \rangle_\beta}{L}.
\end{equation}

For energy transport in the XXZ spin chain, this is especially simple as the energy current $J_E$ is a conserved quantity ($[H,J_E]=0$), so that $\sigma(\omega) = \pi D_E \sigma(\omega)$ with $D_E = \beta \langle J_E^2 \rangle/L$. In this case, the Drude weight can be computed using standard Bethe Ansatz techniques~\cite{0305-4470-35-9-307,2016arXiv160408434Z}. In general, the Drude weight for a current is bounded from below by a sum over conserved quantities that have a nonzero projection on to the current, via the \textit{Mazur inequality}~\cite{Mazur1969533,PhysRevB.55.11029}
\begin{equation}
	\label{quenches_MazurIneq}
	\lim_{t\to\infty} \langle J(t)J(0)\rangle  \geq \sum_{\alpha}\frac{\langle J Q_{\alpha}\rangle^{2}}{\langle Q_{\alpha}Q_{\alpha}\rangle},
\end{equation}
where $Q_{\alpha}$ are independent local conserved quantities~\cite{SUZUKI1971277}.  In the XXZ spin chain at zero magnetic field, the conventional (strictly local) conserved quantities give zero contribution to the Drude weight by symmetry~\cite{sirker:2010}, but a new set of conserved quantities~\cite{prosenxxz} (see also~\cite{PhysRevLett.111.057203,Prosen20141177,1742-5468-2014-9-P09037}) that are given by sums of quasi-local operators (local up to exponential tails) do contribute.  At least at high temperatures and some values of anisotropy $\Delta$, these new integrals of motion appear to saturate the numerical value of the Drude weight~\cite{karraschdrude} obtained from time-dependent density-matrix renormalization group simulations. Two different thermodynamic Bethe ansatz expressions for the spin Drude weight have been proposed~\cite{PhysRevLett.82.1764,doi:10.1143/JPSJS.74S.181}, yielding contradictory results. These Bethe ansatz results are controversial and they were argued to violate exact results at high temperatures (see Ref.~\onlinecite{PhysRevB.83.035115} and references therein). Going beyond linear response, the description of non-equilibrium spin transport in the XXZ model is described by generalized Gibbs ensembles that include non-standard quasi-local conserved quantities~\cite{2016arXiv161207265D}.

\subsection{Spin Drude weights from hydrodynamics}

\begin{figure}[t!]
\includegraphics[width=1.0\linewidth]{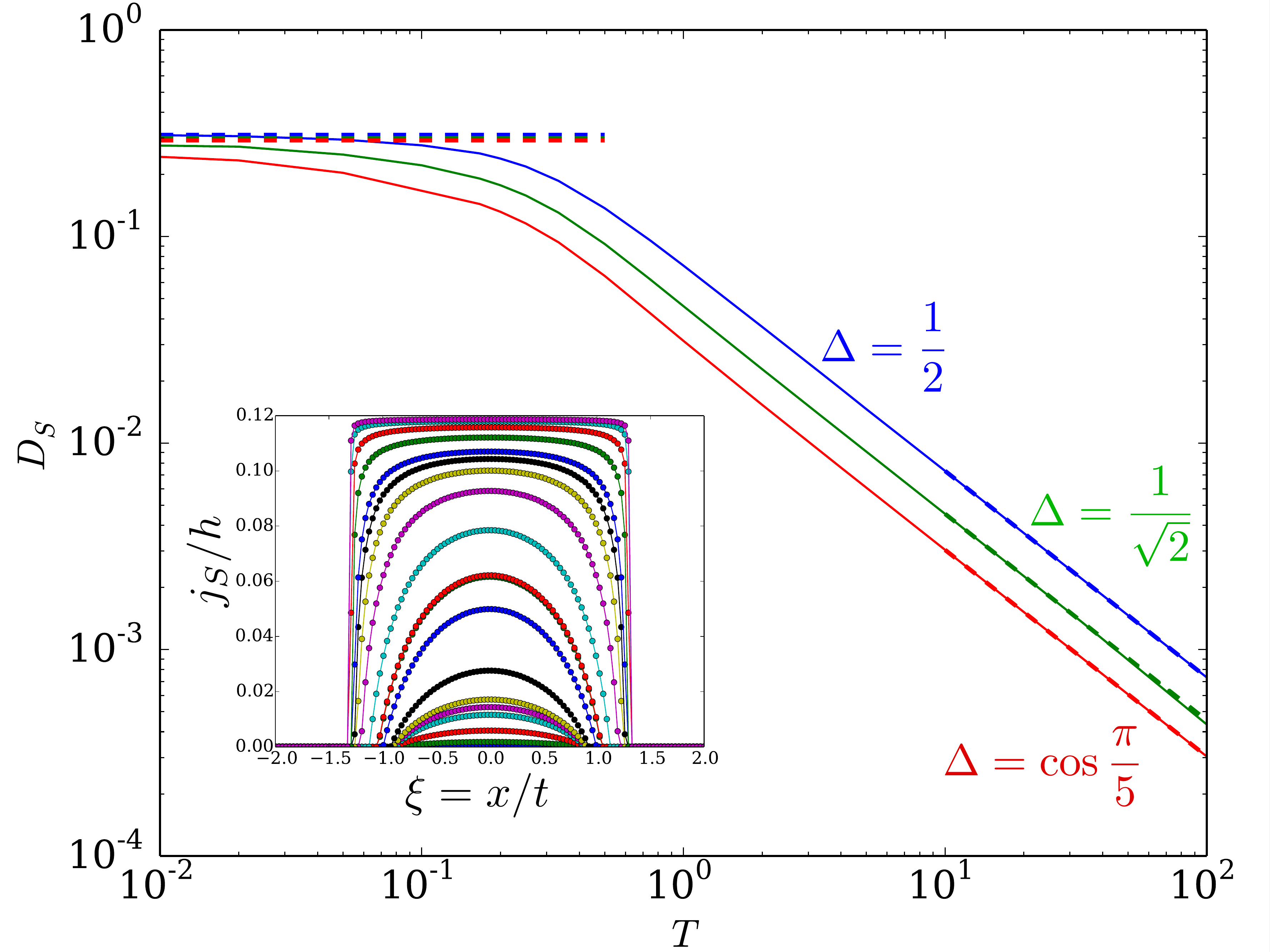}
\caption{Spin Drude weight extracted from the hydrodynamic approach for the XXZ spin chain with $\Delta=\cos \frac{\pi}{\nu}$ and $\nu=3,4,5$. The dashed lines correspond to exact $T=0$ and $T=\infty$ limits (the high temperature result is a lower bound that is believed to be saturated). Inset: steady-state spin current for $h_L=-h_R=h/2=10^{-4}$ and $\Delta=\frac{1}{2}$ for various temperatures, ranging from $T=0.01$ to $T=100$. }
\label{figSpinDrudeHydro}
\end{figure}

Given the long history of linear-response spin transport in the XXZ spin chain, it seems intriguing that the hydrodynamic approach introduced above could describe non-equilibrium spin transport exactly.  In particular, it is interesting to note that the hydrodynamic approach discussed above could have been discovered, in principle, shortly after the development of the TBA formalism~\cite{YY} --- years before the subtleties related to quasi-local charges in the XXZ chain were uncovered by Prosen~\cite{prosenxxz}.
Nevertheless, we will argue below that the Bethe-Boltzmann equation~\eqref{eqBB} can be used to compute both energy and spin Drude weights in agreement with exact low and high temperature results, and with density-matrix renormalization group (DMRG) calculations. 

We start by considering energy transport for a two-temperature quench with a left reservoir at temperature $T_L$ and a right reservoir at temperature $T_R$, joined together at time $t=0$. For a small temperature difference $T_L = T+\Delta T/2$ and $T_R = T-\Delta T/2$ ($\Delta T \ll T$), it is natural to expect that the energy current in the steady state should be described by linear-response theory. In fact, because the energy current is itself a conserved quantity, one can show that the spatial integral of the energy current at long times is determined by the equilibrium Drude weight even \textit{far from equilibrium}~\cite{VKM} 
\begin{equation}
\lim_{t \to \infty} \frac{1}{t} \int_{- \infty}^{\infty} dx \ j_E(x,t) = \int_{T_R}^{T_L} dT \ D_E (T),
\label{eqNESSenergyDrude}
\end{equation}
for arbitrary $T_L$ and $T_R$. In particular, for a small temperature gradient $\Delta T \ll T$, the Drude weight immediately follows from the value of the energy current in the steady state  $ D_E  = \lim_{\Delta T \to 0} \frac{1}{\Delta T} \int d\xi j_E(\xi=x/t)$. Interestingly, equation~\eqref{eqNESSenergyDrude} is exactly satisfied by the Bethe-Boltzmann hydrodynamic approach: in a way similar to the Lieb-Liniger discussion above (Sec.~\ref{secHydroChargesCurrentsLL}), one can show~\cite{Fagotti} that the expectation value of the local energy current $ \langle j_E(x,t) \rangle$ from the hydrodynamic framework coincides with the local conserved quantity $ \langle q_3(x,t) \rangle$, as it should for the XXZ spin chain where the energy current is a conserved quantity. This is a non-trivial check on the hydrodynamic approach. Integrating spatially the conservation law $\partial_t q_3 + \partial_x G=0$, with $G$ the current associated with the conserved charge $q_3$, thus yields $\int dx\,j_E / t = G(T_L)-G(T_R)$, where the ``state function'' $G$ can be determined for a small temperature gradient from linear response~\cite{VKM} . One finds $G(T) = \int^T dT D_E(T)$, from which Eq.~\eqref{eqNESSenergyDrude} follows. 

\begin{figure}[t!]
\includegraphics[width=1.0\linewidth]{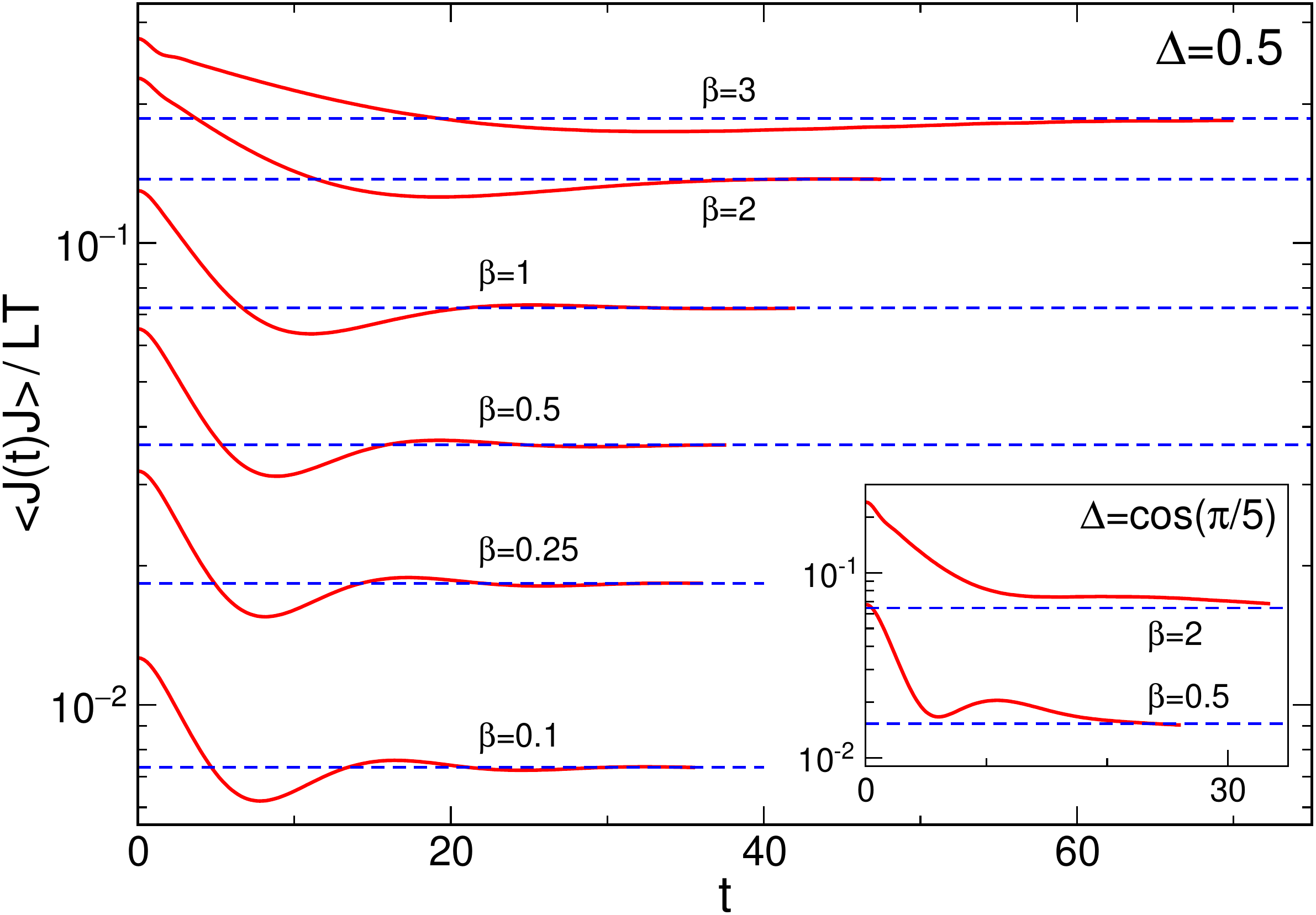}
\caption{Dynamical spin current correlation functions obtained from equilibrium DMRG calculations in the XXZ spin chain with $\Delta=\frac{1}{2}$ and $\Delta=\cos(\pi/5)$ for various temperatures. The long-time asymptotics determine the spin Drude weight $D_s = \beta \lim_{t \to \infty} \frac{\langle J(t) J(0) \rangle_\beta}{L}$ in very good agreement with the Bethe-Boltzmann hydrodynamic approach (dashed lines). }
\label{figSpinDrudeDMRG}
\end{figure}

The hydrodynamic approach is therefore consistent with the linear-response energy Drude weight. This is perhaps not especially surprising, given that the energy Drude weight has a very simple equilibrium expression that is accessible from quantum transfer matrix or thermodynamic Bethe ansatz~\cite{0305-4470-35-9-307,2016arXiv160408434Z}. The case of spin transport between two reservoirs prepared at different magnetic fields $h_L$, $h_R$ and uniform temperature $\beta^{-1}$ is much more interesting. In this case, there is no simple relation like~\eqref{eqNESSenergyDrude} relating non-equilibrium transport and equilibrium Drude weights, but for small fields $h_L=-h_R = \frac{h}{2} \ll J=1$, the spin Drude weight can be expressed as (see Refs.~\onlinecite{VKM,2016arXiv161100573K})
\begin{equation}
D_S=\lim_{h \to 0}\lim_{t \to \infty} \frac{1}{h t} \int_{- \infty}^{\infty} dx \, j_S(x,t),
\label{eqNESSspinDrude}
\end{equation}
which is a spatial integral over the steady-state spin current. This formula allows us to extract the spin Drude weight by solving~\eqref{eqNum} iteratively, and using the hydrodynamic formula
\begin{equation}
j_S(\zeta) = \frac{1}{2} \sum_j \int d\lambda \, n_j\rho_j(\zeta,\lambda)v_j(\zeta,\lambda),
\end{equation}
for the steady-state spin current in a two-reservoir quench with $T_L = T_R$ and $h_R = -h_L = \frac{h}{2} \ll 1$.

\begin{figure}[t!]
\includegraphics[width=1.0\linewidth]{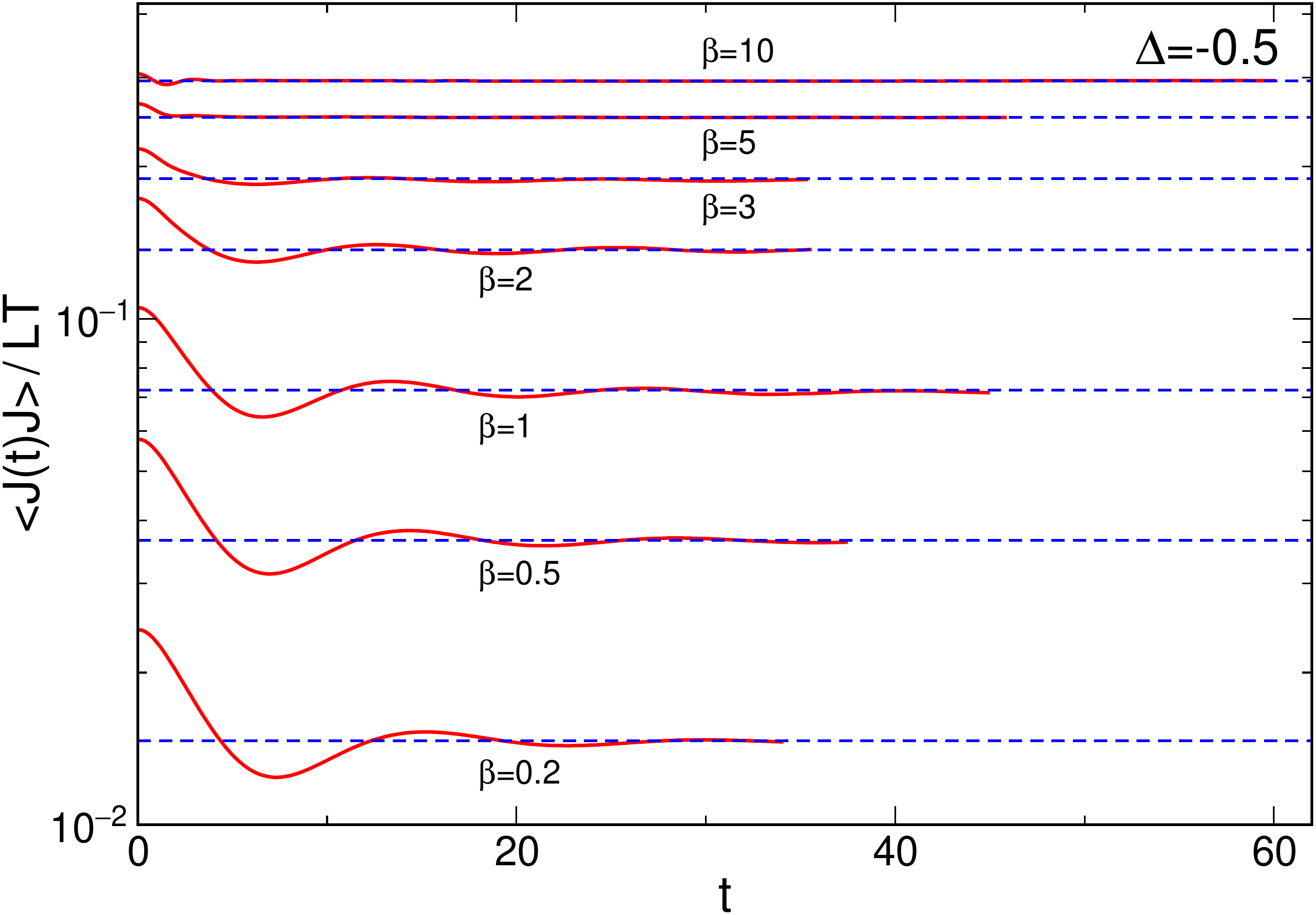}
\caption{Dynamical spin current correlation functions obtained from equilibrium DMRG calculations in the XXZ spin chain with $\Delta=-\frac{1}{2}$, compared with Drude weights from hydrodynamics (dashed lines) for various temperatures. Although the results obtained for the Drude weights differ from the case $\Delta = \frac{1}{2}$, there is clear long-time agreement between DMRG and hydrodynamic predictions, and the DMRG is observed to relax to the latter values even at low $T$.}
\label{figDrudeFerro}
\end{figure}

Results are shown in Fig.~\ref{figSpinDrudeHydro} for different values of the XXZ anisotropy parameter $\Delta$. The small field gradient was taken to be either $h=10^{-3}$ or $h=10^{-4}$ with no significant difference in the value of the Drude weight, and we carefully checked for $\Delta=\frac{1}{2}$ that all the numerical errors associated with the numerical discretization of the hydrodynamic equations lead to relative errors on the Drude weight below $10^{-4}$. Our results are in good agreement with exact asymptotic results at low~\cite{PhysRevLett.65.243} and high~\cite{PhysRevLett.111.057203} temperature -- note that the high temperature results are strictly speaking a lower bound that is believed to be saturated for the values of $\Delta$ that we consider~\cite{karraschdrude,PhysRevLett.111.057203}. The speed of convergence of the spin Drude weight to the $T=0$ result decreases as $\Delta$ approaches $\Delta=1$, consistent with earlier results~\cite{PhysRevLett.82.1764}. 

We further checked these results against density-matrix renormalization group (DMRG) calculations. To this end, we used a finite-temperature version \cite{karraschdrude,barthel13,kennes16} of the real-time DMRG \cite{white92,schollwoeck} to compute the linear response current-correlation function $\langle J(t)J(0)\rangle_\beta$ that appears in Eq.~(\ref{drude_lr}). The key parameter governing the accuracy of this method is the so-called discarded weight, which we chose such that the error of $\langle J(t)J(0)\rangle_\beta$ was smaller than the linewidth. The system size was taken large enough for all results to be effectively in the thermodynamic limit (a typical value is $L\sim200$).

The results of the hydrodynamic approach turn out to match the DMRG predictions to within numerical accuracy (see Fig.~\ref{figSpinDrudeDMRG} and Appendix~\ref{SecAppDMRG}), except at low temperatures where extracting the Drude weight from DMRG calculations would require accessing longer time scales. Thus it seems that the non-equilibrium steady-state predicted by hydrodynamics is cognizant of the quasi-local conserved charges discovered in Refs.~\onlinecite{prosenxxz,PhysRevLett.111.057203,Prosen20141177,1742-5468-2014-9-P09037} (see also~\onlinecite{2016arXiv161207265D}).

It is also instructive to discuss the Drude weight for negative values of $\Delta$, for which bosonization\cite{PhysRevB.83.035115} predicts a small decay rate at low temperatures and finite-time data need to be interpreted with caution. In this regime, hydrodynamics can be viewed as a check on results obtained from the DMRG. In Fig.~\ref{figDrudeFerro}, we show that the DMRG curves saturate to the hydrodynamics prediction even at low $T$. This shows that almost all of the spectral weight is concentrated in the Drude peak and that the DMRG data for the Drude weight \cite{PhysRevB.87.245128} is not misleading.

\section{Discussion and a strongly non-equilibrium example}
\label{SecIV}
\begin{figure}[t!]
\includegraphics[width=1.0\linewidth]{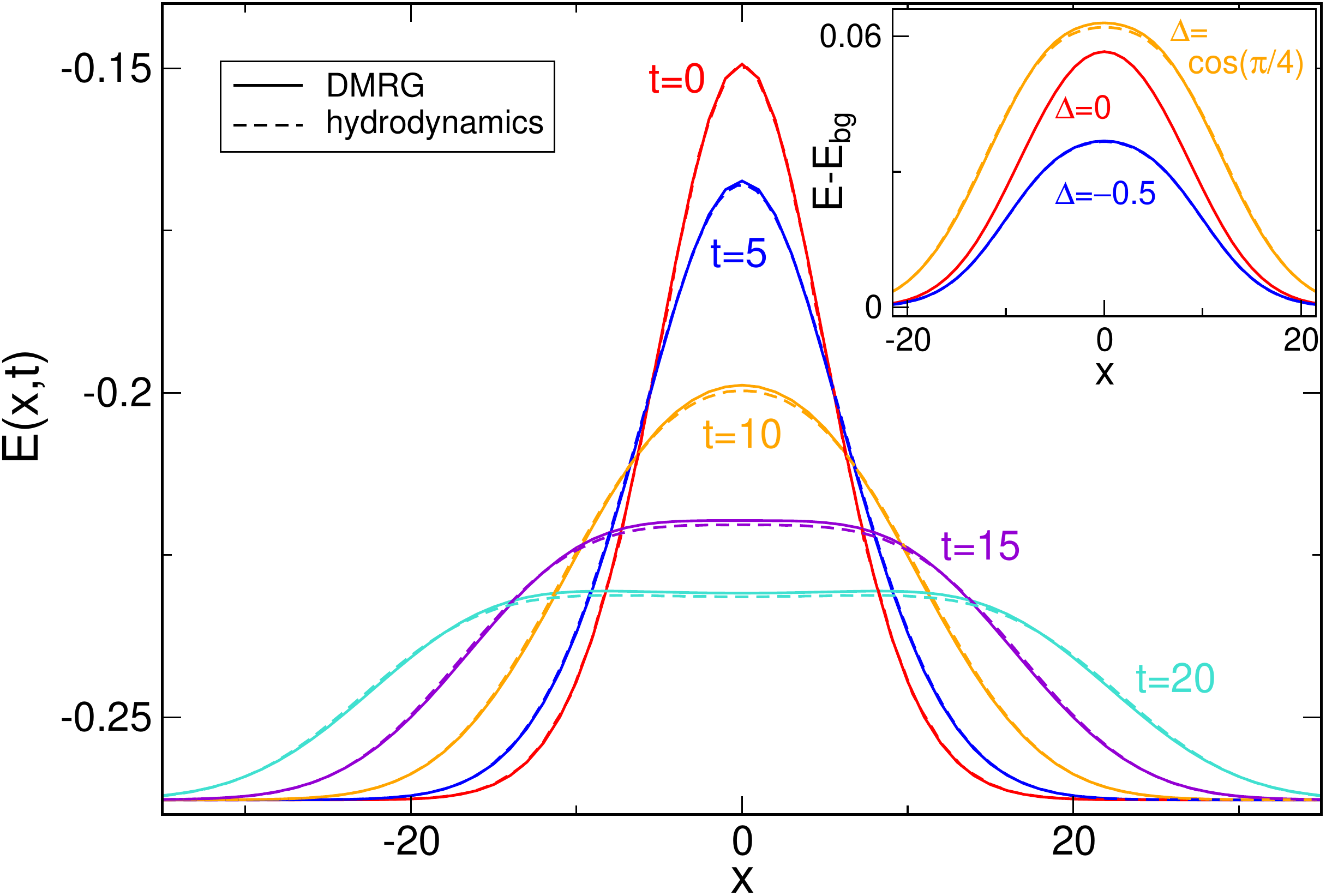}
\caption{Expansion in time of a Gaussian temperature profile $\beta(x) = \beta_0 - (\beta_0-\beta_M) {\rm e}^{-x^2 / L^2}$ in the XXZ spin chain with $\Delta=-\frac{1}{2}$, $\beta_0=2$, $\beta_M=1$ and $L=8$. The agreement between the one-step implicit hydrodynamic solution~\eqref{scinf} (dashed lines) and DMRG results (solid lines) is excellent. Inset: finite time solution at $t=10$ for different values of the XXZ anisotropy parameter $\Delta$.}
\label{figFiniteTime}
\end{figure}

Having treated linear response in the XXZ chain in the previous section, we now consider the following far-from-equilibrium example of the implementation described in Section \ref{SecII}.  Thus suppose that a system is prepared with a Gaussian temperature profile
\begin{equation}
\beta(x) = \beta_0 - (\beta_0-\beta_M) {\rm e}^{-x^2 /L^2},
\label{eqGaussianProfile}
\end{equation}
where $L$ characterizes the spread of the initial localized temperature inhomogeneity. We compared the results of the one-step implicit hydrodynamic solution~\eqref{scinf} to DMRG simulations starting from the same initial profile~\eqref{eqGaussianProfile} for various values of $\beta_0$, $\beta_M$ and $\Delta$ and found an excellent agreement (Fig.~\ref{figFiniteTime}). 

The agreement is perhaps surprising given the way that Eq.~\eqref{scinf} is motivated in Appendix \ref{SecAppC}, as an implementation of a finite-difference scheme with only one time step. We do not yet understand this level of agreement. It could arise because there are a number of limits or simple cases for which the finite-difference scheme is actually exact, which constrains the behavior even in other cases such as the Gaussian initial profile. It is also possible that the scheme is accurate to higher order. We will discuss more examples and mathematical aspects of this approach in a future publication~\cite{Unpub}.

A number of natural questions arise for future work. The first concerns the limitations of the hydrodynamical approach. For example, the Bethe-Boltzmann equation is effectively classical in the sense that nearly the same mathematical structure would arise in the dynamics of an integrable classical particle gas. (For a recent study of the two-reservoir quench in classical systems and references to earlier classical work, see Ref.~\onlinecite{spohn}.) Where did quantum-mechanical effects, reflecting the wave nature of the particles, go? Obviously the quantum-mechanical interactions between particles determine the phase shifts that underlie the Bethe ansatz, but it is apparently true that long-time dynamics in many situations is effectively classical. This includes situations such as the two-reservoir quench that are sufficient to determine the Drude weights. It should be possible to discern quantum effects in short-time or short-length behavior, which could be viewed as transients before the local GGE assumption of generalized hydrodynamics becomes justified. We should also note that there is no reason to expect the striking agreement in Fig. \ref{figFiniteTime} for observables such as spin density, which has two transport channels (ballistic and diffusive) even in the gapless regime~\cite{PhysRevB.83.035115}. This is in contrast to energy density, whose transport is purely ballistic across all regimes, and is related to the fact that the energy current is a conserved charge of the XXZ Hamiltonian whereas the spin current is not.

Aside from numerical studies, a complementary approach that might be useful for understanding the scope of hydrodynamical techniques is based on exact solutions for time evolution in certain limits, such as the Luttinger-liquid two-reservoir quench studied in recent work~\cite{2017arXiv170106620L}. This should be comparable to the XXZ model studied in this paper in the low-temperature regime. At intermediate times, features are seen in the time evolution of densities that are compared to those in numerical calculations~\cite{karraschilanmoore} and may reflect finite-time corrections to the hydrodynamic description. For the case of energy transport, similar terms, also involving the Schwarzian derivative of the initial temperature distribution, appeared previously in a calculation based on conformal invariance~\cite{sotiriadiscardy}.

Another obvious question concerns the mathematical existence and physical validity of the hydrodynamical solutions in Section \ref{SecIIC}, beyond the special case of the two-reservoir quench. The two-reservoir quench is quite special for a number of reasons: for example, it has no intrinsic time or length scale, which means that the scaling limit is effectively a complete description of its universal properties. We have found that for at least some cases of practical importance, computations based on the implicit method described here yield physically plausible results, even after the time at which shocks from different initial discontinuities coincide. However, this is very far from a mathematical demonstration of existence of solutions, which may be unreasonable to expect given that no such proof exists even for the venerable equations of standard hydrodynamics.

A deeper physical understanding of the behavior of the Bethe-Boltzmann equation in various important contexts is a more feasible goal. Even for the two-reservoir quench, important questions remain. We have limited ourselves in this paper to the regime $-1 < \Delta < 1$, when the dynamics have a ballistic component (e.g., the energy and spin Drude weights are nonzero). A very recent numerical study~\cite{znidaric} finds diffusive behavior for $|\Delta| > 1$ (see also Refs.~\onlinecite{PhysRevB.79.214409,PhysRevLett.107.250602,PhysRevLett.106.220601,PhysRevB.89.075139}) and superdiffusive behavior at the Heisenberg points $|\Delta| = 1$, and it would be desirable yet difficult to capture this behavior using the hydrodynamic formalism. While it is no doubt challenging to capture the entire diversity of dynamical behavior in integrable models within a single formalism, the Bethe-Boltzmann equation, or equivalently generalized hydrodynamics, is at an exciting stage of its development, with important results for some long-standing problems and tantalizing hints for others.

\smallskip

{\it Acknowledgments.}
We thank Fabian Essler for several valuable discussions. We also wish to thank J. De Nardis, B. Doyon, M. Fagotti, and E. Ilievski for comments on the manuscript. J.E.M. and V.B.B. are grateful to the Institut d'{\'E}tudes Scientifiques de Carg{\`e}se and the organizers of the conference ``Exact Methods in Low-Dimensional Statistical Physics 2017'' for their hospitality. We acknowledge support from the Department of Energy through the LDRD program of LBNL (R.V.), from NSF DMR-1507141 and a Simons Investigatorship (J.E.M.), and center support from the Moore Foundation's EPiQS initiative. C.K. is supported by the DFG via the Emmy-Noether program under KA 3360/2-1.

{\it Note added.}
While this work was being completed, we became aware of a related paper~\cite{2017arXiv170202930I}, which also showed that the spin Drude weight could be obtained from hydrodynamics. 

\bibliography{NESS}

\eject

\

\newpage

\onecolumngrid

\appendix
\section{Spin Drude weight from DMRG}
\label{SecAppDMRG}
In the main text, we compared DMRG calculations of the spin Drude weight to the hydrodynamic approach for $\Delta=\frac{1}{2}$. We also performed detailed comparisons for $\Delta=\frac{1}{\sqrt{2}}$ (left) and $\Delta=\cos \frac{\pi}{5}$ with good agreement, even though the convergence of the DMRG calculations to the asymptotic values becomes slower as $\Delta$ gets closer to the isotropic value $\Delta=1$.\\
\begin{figure*}[h]
\includegraphics[width=0.48\linewidth]{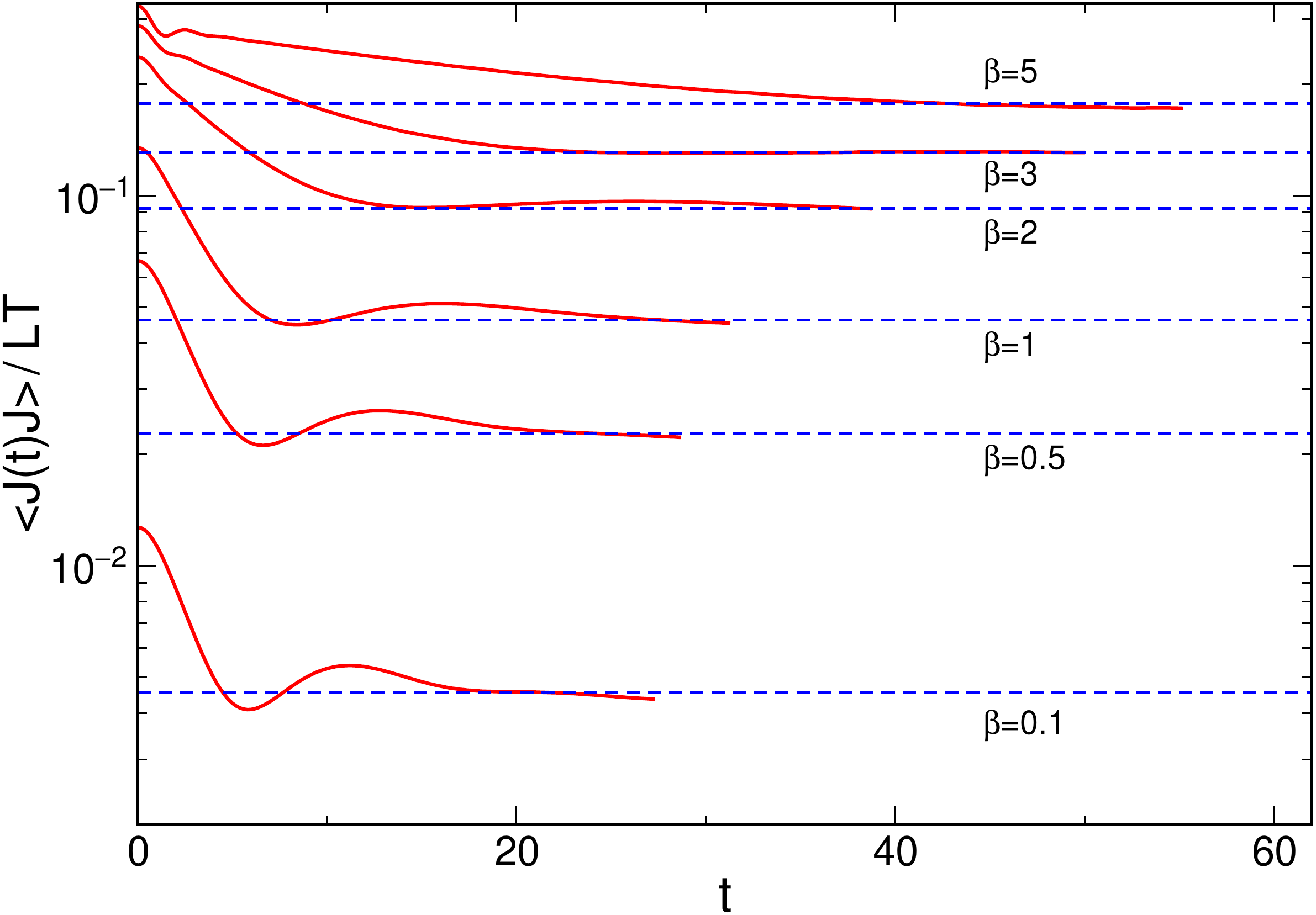}\hspace*{0.02\linewidth}
\includegraphics[width=0.48\linewidth]{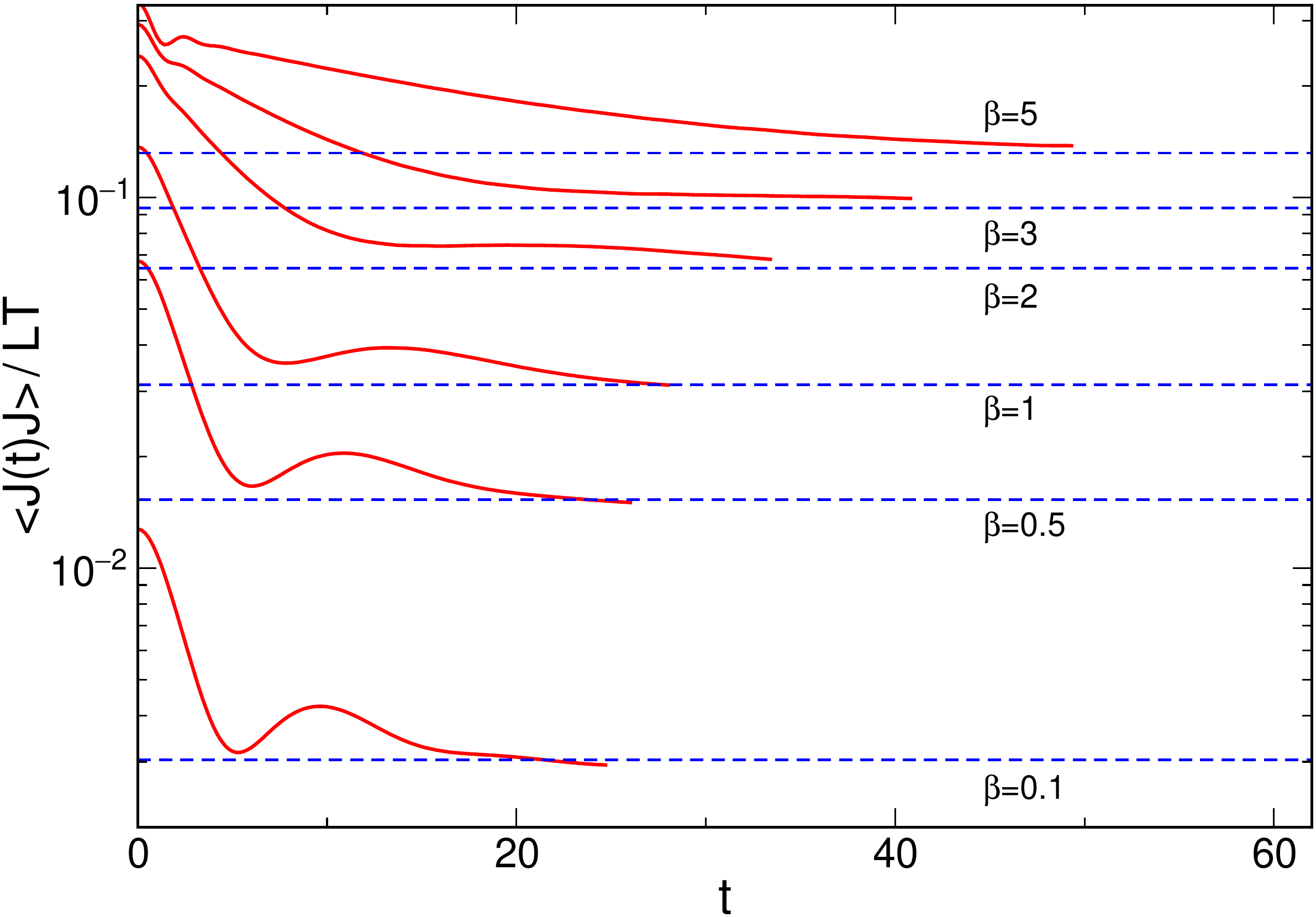}
\caption{Dynamical spin current correlation functions obtained from equilibrium DMRG calculations in the XXZ spin chain with $\Delta=\frac{1}{\sqrt{2}}$ (left) and $\Delta=\cos \frac{\pi}{5}$ (right) for various temperatures. The results for the spin Drude weight at long times are in good agreement with the hydrodynamic predictions (dashed lines).}
\label{figSpinDrudeDMRG2}
\end{figure*}

\section{Summary of Thermodynamic Bethe Ansatz}
\label{AppB}
\subsection{Lieb-Liniger Gas}
In this section, we briefly recall the main features of thermodynamic Bethe ansatz for the Lieb-Liniger gas. Recall that the Lieb-Liniger Hamiltonian for $N$ bosons on a line, with interaction strength $c$, may be written as
\begin{equation}
\hat{H} = - \sum_{j=1}^N \partial_j^2 + \sum_{j<k}2c\delta(x_j-x_k).
\end{equation}
Away from the collision planes $\{x_j=x_k : j\neq k\}$, the eigenfunctions of this Hamiltonian are superpositions of plane wavefunctions. For example, in the fundamental chamber $\mathcal{D}_0 = \{x_1<x_2<\ldots<x_N\}$ we may write
\begin{equation}
\Psi(x_1,x_2,\ldots,x_N) = \sum_{P \in S_N} A(P) e^{i\sum_jk_{P(j)}x_j},
\end{equation}
where the sum is over all permutations on $N$ letters. The model is called \textit{integrable} because all $N$-body scattering processes factorize into two-body processes. This property turns out to be sufficiently restrictive to solve the model. In particular, upon imposing periodic boundary conditions
\begin{equation}
\Psi(x_1,x_2,\ldots,x_N) =\Psi(x_2,x_3,\ldots,x_N,x_1+L),
\end{equation}
one may use certain combinatorial relations between the amplitudes $A(P)$ to deduce the \textit{Bethe equations}
\begin{equation}
k_aL = 2\pi I_a + \sum_{b\neq a}\varphi(k_a-k_b), \quad a = 1,2,\ldots, N,
\label{Bethe}
\end{equation}
where $\varphi(k)$ denotes the two-body phase shift. The \textit{Bethe quantum numbers} $I_a$ are integers for fermions or an odd number of bosons, and half-integers for an even number of bosons. Generally one finds that only certain allowed Bethe quantum numbers are occupied in a given eigenstate of the Hamiltonian. Passing to the thermodynamic limit $N,L\to \infty$, with $N/L$ fixed, this allows one to define densities of \textit{particles}, \textit{holes} and \textit{vacancies} via
\begin{align}
\nonumber L \rho(k) dk &= \{ \textrm{\# occupied wavenumbers in $[k,k+dk)$} \} \\
\nonumber L \rho^h(k) dk &= \{ \textrm{\# unoccupied wavenumbers in $[k,k+dk)$}\} \\
L \rho^t(k) dk &= \{ \textrm{\# allowed wavenumbers in $[k,k+dk)$}\}.
\end{align}
respectively. By definition, we have
\begin{equation}
\rho^t(k) = \rho(k) + \rho^h(k),
\end{equation}
and it is useful to define the \textit{non-equilibrium Fermi factor}, given by
\begin{equation}
\vartheta(k) \defeq \frac{\rho(k)}{\rho^t(k)}.
\end{equation}
One can then show~\cite{Takahashi} that the continuum version of the Bethe equations ~\eqref{Bethe} reads
\begin{equation}
\rho^t(k) + \int_{-\infty}^{\infty} dk'\, \mathcal{K}(k,k')\vartheta(k')\rho^t(k') = \frac{1}{2\pi},
\label{TBA}
\end{equation}
where the kernel $\mathcal{K}$ is given by
\begin{equation}
\mathcal{K}(k,k') =\frac{1}{2\pi}\varphi'(k-k') =  -\frac{c}{\pi}\frac{1}{c^2 + (k-k')^2}.
\end{equation}
Integral equations of this type occur sufficiently frequently in the non-equilibrium theory that we define operators $\hat{K}$ and $\hat{\vartheta}$ which act on functions via
\begin{align}
\hat{K}[f](k) &= \int_{-\infty}^{\infty} dk'\,\mathcal{K}(k,k')f(k'), \\
\quad \hat{\vartheta}[f](k) &= \vartheta(k)f(k).
\end{align}
For example, in operator notation, equation ~\eqref{TBA} reads
\begin{equation}
(\hat{1} + \hat{K}\hat{\vartheta})[\rho^t](k) = \frac{1}{2\pi}. 
\end{equation}
Now consider particle-type excitations on such states. These arise when one introduces an additional quantum number in the Bethe equations \eqref{Bethe} causing a shift in wavenumber $k_j \to k_j'$ across all $k$-space, which reflects the collective nature of the underlying excitation. The physics of such excitations is captured by the pair of integral equations,
\begin{align}
\nonumber (\hat{1} + \hat{K}\hat{\vartheta})[\partial_k F](k'|k) &= -\mathcal{K}(k,k'), \\
Q'(k) + \int_{-\infty}^{\infty} dk'\,\partial_kF(k'|k)\vartheta(k')Q'(k') &= \Delta Q'(k),
\label{backflow}
\end{align}
where $F(k'|k) \defeq L \rho^t(k)\Delta k$ denotes the \textit{backflow function}\footnote{Our convention for the backflow function follows Ref.~\onlinecite{Fagotti}, and in that respect is different from the convention $F(k'|k) = L \rho(k)\Delta k$ used, for example, in Ref.~\onlinecite{Takahashi}.} for a particle-type excitation with momentum $k$, and $Q$ and $\Delta Q$ denote the bare and dressed values respectively for any given conserved charge of the model. Rather conveniently, the equations~\eqref{backflow} together imply the closed integral equation
\begin{equation}
\Delta Q'(k) + \int_{-\infty}^{\infty} dk'\,\mathcal{K}(k,k')\vartheta(k')\Delta Q'(k') = Q'(k)
\label{dc}
\end{equation}
for the dressed charge. In the context of equilibrium TBA, this equation is used to justify the interpretation of $\vartheta$ as a Fermi factor~\cite{Korepin,Takahashi}. In the present context, it allows us to determine the dressed charges carried by a quasiparticle excitation directly from the bare charges. Now recall that the group velocity of a quasiparticle excitation is given by
\begin{equation}
v(k) = \frac{\Delta E'(k)}{\Delta P'(k)},
\end{equation}
where $\Delta E$ and $\Delta P$ denote its dressed energy and momentum respectively. From the bare values $P(k) = k, \, E(k) = k^2/2$, we can take the formal inverse\footnote{This is shorthand for the infinite (Neumann) series obtained by solving ~\eqref{dc} iteratively.} of the kernel appearing in ~\eqref{dc} to yield the dressed values
\begin{align}
\nonumber \Delta P'(k) &=  (1+\hat{K}\hat{\theta})^{-1}[1](k), \\
\Delta E'(k) &=  (1+\hat{K}\hat{\theta})^{-1}[k'](k),
\end{align}
so that the quasiparticle velocity is given by
\begin{equation}
v(k) = \frac{(\hat{1}+\hat{K}\hat{\vartheta})^{-1}[k'](k)}{(\hat{1}+\hat{K}\hat{\vartheta})^{-1}[1](k)}.
\end{equation}

\subsection{The Gapless XXZ Chain}
Recall that the Hamiltonian for the spin-$1/2$ XXZ chain on $N$ sites, in zero external field, is given by
\begin{equation}
H = J \sum_{j=1}^{N-1} S^x_{j}S^x_{j+1}+S^y_{j}S^y_{j+1} + \Delta S^z_{j}S^z_{j+1}.
\end{equation}
We take periodic boundary conditions $S_N \equiv S_{N+1}$, set the coupling to $J=1$, and parameterize the anisotropy of the theory by $\Delta = \cos \gamma$.  We also assume that the model is in its gapless phase, i.e. $-1 < \Delta < 1$. The thermodynamic limit of this phase is believed to be characterized by a finite number of ``strings'' of bound states, consisting of families of rapidities in the complex plane possessing the same real part~\cite{Takahashi}. We define a \textit{string of type $j$} to be a an ordered pair, $(n_j,v_j)$, where $n_j$ is the number of spin-flips comprising the string and $v_j$ is its parity. Suppose that there are $N_t$ string types in total so that $j \in \{1,2,\ldots,N_t\}$, and $M_j$ strings of type $j$. Let $M$ denote the total number of spin-flips in the system. By definition, $\sum_{j=1}^{N_t} M_jn_j = M$. Upon fixing a string type $j$, we denote the rapidities of a given string $\alpha \in \{1,2,\ldots,M_j\}$ within that type by 
\begin{equation}
\lambda^j_{\alpha,a} = \lambda_{\alpha}^{j} + i[(m_j+1-2a)\gamma+ (1-v_j)\frac{\pi}{2}], \quad a = 1,2,\ldots,m_j\end{equation}
The TBA equations then read
\begin{equation}
\sigma_j[\rho_j(\lambda)+ \rho^h_j(\lambda)] + \sum_{k=1}^{N_t}\int_{-\infty}^{\infty}d\lambda' \, T_{jk}(\lambda-\lambda')\rho_k(\lambda')= a_j(\lambda),
\end{equation}
where $j \in \{1,2,\ldots,N_t\}$; for definitions of the various terms see the book of Takahashi~\cite{Takahashi}. We define quantities
\begin{equation}
\eta_j(\lambda) = \frac{\rho^h_j(\lambda)}{\rho_j(\lambda)},
\end{equation}
and the Fermi factors
\begin{equation}
\vartheta_j(\lambda) = \frac{\rho_j(\lambda)}{\rho^t_j(\lambda)} = \frac{1}{1+\eta_j(\lambda)},
\end{equation}
for strings of type $j$. It can be shown~\cite{Fagotti} that the dressed charge for any given quasiparticle excitation is related to the bare charge via the $N_t$ coupled integral equations
\begin{equation}
\Delta Q'_j(\lambda) + \sum_{k=1}^{N_t}\int_{-\infty}^{\infty}d\lambda' \, T_{jk}(\lambda-\lambda')\vartheta_k(\lambda') \sigma_k\Delta Q'_k(\lambda') = Q_j'(\lambda').
\end{equation}
These imply the formula
\begin{equation}
v_j(\lambda) =  \frac{1}{2\pi}\frac{(\hat{\sigma}+\hat{T}\hat{\vec{\vartheta}})^{-1}[-A\vec{a}']_j}{(\hat{\sigma}+\hat{T}\hat{\vec{\vartheta}})^{-1}[\vec{a}]_j}(\lambda)
\end{equation}
for the velocity of quasiparticle excitations within each string type.

\section{Finite-time Numerical Scheme}
\label{SecAppC}
Consider the infinite-dimensional initial value problem 
\begin{align}
\nonumber \partial_t \vartheta(x,t,k) + v[\hat{\vartheta}](k)\partial_x \vartheta(x,t,k) &= 0, \\
\vartheta(x,0,k) &= \phi(x,k).
\label{IVP}
\end{align}
Motivated by analogy with the solution by characteristics for the Burgers equation, we propose the compact approximate ansatz
\begin{equation}
\vartheta(x,t,k) = \phi(x-v[\hat{\vartheta}](k)t,k).
\label{scinf}
\end{equation}
Here the velocity functional $v[\hat{\vartheta}]$ is evaluated using all values of $\vartheta$ at $(x,t)$.  This ansatz is presumably not exact in general but arises naturally as the one-step version of a backwards implicit scheme to solve the advection equation Eq.~\eqref{IVP}.~\footnote{We thank B.~Doyon for helpful discussions and for sharing unpublished work on a different approach to this problem.}  We solve it numerically for an example in the closing discussion section of this paper (see Fig. \ref{figFiniteTime} above).

First, in order to provide some intuition, we illustrate the meaning of this ansatz for the finite-dimensional version of Eq.~\eqref{IVP} (as arises in practice when discretizing on a computer). Thus consider the initial-value problem
\begin{align}
\nonumber \partial_t \theta_n(x,t) + v_n(\theta_1,\theta_2,\ldots,\theta_N)\partial_x \theta_n(x,t) &= 0 \\
\theta_n(x,0) &=\phi_n(x),
\label{dIVP}
\end{align}
for $t>0$, with $n=1,2,\ldots, N$.  We will have found a solution of this equation (ignoring the possibility of discontinuities in any derivative) if we can find $\theta_n$ satisfying
\begin{align}
&\theta_n(x_0+ dt\,v_n(\theta_1,\theta_2,\ldots,\theta_N)|_{x_0,t_0},t_0+dt) \notag
\\ &= \theta_n(x_0,t_0) + O(dt^2),\label{dt2}
\end{align}
for every $x_0,t_0$ and small $dt$, because expanding the left-hand side in a Taylor series yields the advection equation.  Since the error is of order $dt^2$, this could be the basis of a convergent finite-difference scheme: given $\theta_n$ at time $t_0$, one steps forward in time repeatedly by $dt$ using Eq.~\eqref{dt2}, and the global error in advancing by time $t$ is small as $dt \rightarrow 0$.
However, it is well-known from the theory of simpler differential equations that this type of Euler method can be quite unstable.  A safer option is to use an implicit or backwards method~\cite{numrep}: for a time step $dt$, we solve for the current values of $\theta_n$ by looking back at earlier times.
\begin{equation}
\theta_n(x,t) = \theta_n(x-v_n(\theta_1,\theta_2,\ldots,\theta_N)|_{x,t}\,dt,t-dt),
\label{scf}
\end{equation}
where the notation means that the velocities are evaluated at ${x,t}$.  This is just the infinitesimal form of Eq.~\eqref{scinf}; in other words, Eq.~\eqref{scinf} is obtained by covering the entire desired integration range of time in a single step.  Another advantage in practice of the backwards scheme is that, unlike for the forward scheme, it is easy to get all the $\theta_n$ at a particular $(x,t)$ point without interpolation if the initial condition is known everywhere.

We have found that one can solve the ansatz~\eqref{scf} by numerical iteration to obtain detailed predictions for finite-time evolution, from a range of initial conditions~\cite{Unpub}. This is of particular experimental relevance for the Lieb-Liniger model, which provides a physically realistic description of quasi one-dimensional Bose gases. We report one such calculation for energy expansion in the XXZ chain at the end of this paper using the simpler approximate ansatz~\eqref{scinf} (corresponding to a single time step with $dt=t$ in eq.~\eqref{scf}), after introducing that model and discussing its linear response, and discuss there its (perhaps unexpected) agreement with a DMRG solution. A detailed discussion of the range of validity of the predictions obtained using this method is given in Sections \ref{SecII} and \ref{SecIV} of the main text.


\end{document}